\documentstyle[mprocl]{article}

\def\be{\begin{equation}}
\def\ee{\end{equation}}
\def\bea{\begin{eqnarray}}
\def\eea{\end{eqnarray}}

\begin{document}

\title{Semiclassical Quantization on Black Hole Spacetimes}

\author{ R. Casadio, B. Harms and Y. Leblanc }

\address{Department of Physics and Astronomy, The University
of Alabama,\\ Box 87034 Tuscaloosa, AL 35487-0324 USA}

\maketitle\abstracts{The microcanonical treatment of black holes as opposed to the canonical         
formulation is reviewed and some major differences are displayed.               
In particular the black hole decay rates are 
compared in the two different pictures.}

The treatment of black holes as thermodynamical systems has many                
mathematical and physical inconsistencies and drawbacks.                        
In this approach the specific heat                         
turns out to be negative, which is a clear signal       
that the thermodynamical analogy fails.                                         
A second problem is that the partition function as calculated                    
from the microcanonical density of states                                      
is infinite \cite{hl1} for all temperatures and hence the canonical ensemble               
is not equivalent to the (more fundamental) microcanonical ensemble,            
as is required for thermodynamical equilibrium.                                 
\par                                                                            
Black        
holes can be shown to radiate \cite{hawk} and in the thermodynamical            
approach the radiation coming out of the black hole has a Planckian             
distribution.                                                                    
Since black holes can in principle radiate away completely,                     
this result implies that information can be lost, because                       
pure states can come into the black hole but only mixed states                  
come out.                                                                       
The breakdown of the unitarity principle is one of the most serious             
drawbacks of the thermodynamical interpretation, since it requires              
the replacement of quantum mechanics with some new (unspecified)                
physics.                                                                        
\par                                                                            
                                   
We have investigated \cite{hl1}-\cite{mfdbh} an alternative description of black holes which            
is free of the problems encountered in the thermodynamical approach.            
In our approach black holes are considered to be extended quantum objects       
($p$-branes) with degeneracy proportional to the inverse of the                 
probability to tunnel through the black hole's horizon,                         
$\sigma \simeq c\,e^{A/4}$.                 
Explicit expressions for $\sigma$ can be obtained for some                      
geometries, {\em e.g.} the $D$-dimensional Schwarzschild black hole             
with mass $M$,                                                                  
\begin{eqnarray}                                                                
\sigma(M) \sim e^{C(D)\,M^{D-2/D-3}}                                            
\ .                                                                             
\label{D-sc}                                                                    
\end{eqnarray}                                                                  
where $C(D)$ is a mass-independent function.                                    
An exponentially        
rising density of states is the clear signal of a non-local field theory\cite{fub}.       
Comparing Eq.~(\ref{D-sc}) to those known for non-local field theories,         
we find that it corresponds to the degeneracy of states for a $p$-brane         
of dimension $p =(D-2)/(D-4)$.                                                  
\par                                                                            
A second example is the Kerr-Newman dilaton black hole                     
\cite{chlc,chl}                                                                 
An analysis of the microcanonical density          
$\Omega(M,J,Q;V)$ of a gas of such black holes in a volume $V$ shows that the most probable configuration is                      
one massive black hole which carries all the charge and angular momentum        
and is surrounded by $n-1$ lighter, Schwarzschild black holes.                  
This is the statistical mechanical model of a black hole                        
(the most massive one in the system) and its associated radiation               
(whose quanta are represented by the lighter black holes in the gas).           
The bootstrap condition \cite{hag} is satisfied,                    
\begin{equation}                                                                
\lim\limits_{M\to\infty}\,{\Omega(M,Q,J;V,a)\over\sigma_{KND}(M,Q,J;a)}         
=1                                                                              
\ ,                                                                             
\label{boots}                                                                   
\end{equation}                                                                  
\par                                                                            
To study particle production and propagation in black hole geometries we        
apply the mean field approximation where fields are quantized on                
a classical black hole background.                                              
The canonical formulation in which black holes are assumed to have fixed        
temperature $\beta_H^{-1}$ leads to the number density given by the Planckian distribution for Hawking radiation and the corresponding partition        
function violates Hagedorn's self-consistency condition                         
(see \cite{hag,chl} ).                                      
Thus the black hole system is not in thermal equilibrium, and the thermal vacuum $|\beta_H\rangle$ is the false vacuum for a black       
hole.                                                                           
The true vacuum $|E\rangle$ can be obtained by considering black        
holes as having fixed energy $E$.                                               
In the microcanonical ensemble                 
\cite{mfd} the vacuum state is                                                        
\begin{eqnarray}                                                                
|E\rangle = {1\over{\Omega(E)}}\,\int_0^E \Omega(E-E')\,                        
L_{E-E'}^{-1}\left[|\beta_H\rangle\right]\,dE'                                  
\; ,                                                                            
\end{eqnarray}                                                                  
where $L^{-1}$ is the inverse Laplace transform, and the microcanonical      
number density is                                                                 
\begin{eqnarray}                                                                
n_E(\omega) = \sum_{l=1}^\infty\,{\Omega(E-l\,\omega(m))                        
\over\Omega(E)}\,\theta(E-l\,\omega)                                            
\ ,                                                                             
\label{n_E}                                                                     
\end{eqnarray}                                                                  
which is our candidate alternative to the Planckian distribution for particles emitted by a black hole.                              
\par                                                                            
To test the local              
properties of the spacetime which corresponds to the microcanonical vacuum      
$|E\rangle$ we study the propagation of waves as probes .                                                                    
If we now consider waves propagating on a Schwarzschild spacetime, and do       
not take into account back-reactions, the incoming wave becomes            
\begin{eqnarray}                                                                
\psi_{in} = \left\{\matrix{\strut\displaystyle{ {Y_{lm}(\theta,\phi)            
\over{\sqrt{8\,\pi^2\,\omega}}}\,                                               
{e^{i\,(\omega/\kappa)\,\ln(v_0-v)}\over{r}}} &\ \ \ v < v_0\cr                 
\cr                                                                             
0 &\ \ \ v > v_0 \cr}\right.                                                    
\label{psi_T}                                                                   
\end{eqnarray}
where $v = t+r_*$, $u = t-r_*$ and $r_*$ is the {\it tortoise} coordinate.                                                                     
The $in$ states for the two vacua are related by a Bogoliubov                    
transformation with coefficients $\alpha$ and $\beta$                  
\begin{eqnarray}                                                                
|\alpha_{\omega\omega'}|^2 = e^{2\,\pi\,\omega/\kappa}\,                        
|\beta_{\omega\omega'}|^2                                                       
\ .                                                                             
\label{ab}                                                                      
\end{eqnarray}                                                                  
Substituting Eq.~(\ref{ab}) into the Wronskian relation                         
$\sum_{\omega'}[|\alpha_{\omega\omega'}|^2 -|\beta_{\omega\omega'}|^2]          
= 1$ one obtains again the Planckian distribution.        
\par                                                                            
If in Eq.~(\ref{psi_T}) we make the formal replacement \cite{mfdbh}             
${2\,\pi\,\omega\over{\kappa}} \to \ln\left(1+n_E^{-1}(\omega)\right)            
\;$ ,                                                                            
where $n_E(\omega)$ is the microcanonical number density in Eq.~(\ref{n_E}),    
the coefficients in the Bogoliubov transformation                                
satisfy                                                                         
\begin{eqnarray}                                                                
|\alpha_{\omega\omega'}|^2 = e ^{\ln(1+n_E^{-1}(\omega))}\,                     
|\beta_{\omega\omega'}|^2                                                       
\ ,                                                                             
\end{eqnarray}                                                                  
which gives the required number density $n_E$.                                  
The wave so obtained does not solve the same wave                     
equation as the wave in Eq.~(\ref{psi_T}), but it will solve a wave             
equation in a background whose metric includes back-reaction and                
non-local effects.                                                              
\par                                                                            
As an example of the differences between the predictions                        
of the two approaches (thermal vs. microcanonical)                              
we consider 4-dimensional Schwarzschild black holes with                        
microcanonical density $\Omega(M) = e^{4\,\pi\, M^2}$                           
(in this case $E$ = $M$).                    
We then find for the number density                                             
\begin{eqnarray}                                                                
n_M(\omega) = \sum_{l=1}^{M/\omega} e^{-8\,\pi\, M\,l\,\omega +                 
4\,\pi\, l^2\,\omega^2}                                                         
\; .                                                                            
\label{n_M}                                                                     
\end{eqnarray}                                                                  
Using the expressions for $n_{\beta_H}$ and $n_M$      
in Eq.~(\ref{n_M}), we can compare the decay rates for radiating                
black holes predicted by the two theories \cite{mfdbh}.                         
We find for the thermodynamical rate                                            
\begin{eqnarray}                                                                
{dM\over{dt}} \sim -{1\over{M^2}}                                               
\ ,                                                                             
\end{eqnarray}                                                                  
and for the microcanonical rate                                                 
\begin{eqnarray}                                                                
{dM\over{dt}} \sim -M^6                                                         
\ .                                                                             
\end{eqnarray}                                                                  
The thermodynamical theory predicts that it takes a finite            
time for the black hole to completely evaporate, thus precluding the              
presence of primordial black holes in our Universe.                             
The microcanonical rate, on the other hand, predicts that primordial,           
microscopic black holes could still be around today, since the                  
decay rate goes to zero as a power of $M$.                                      

\section*{Acknowledgments}
This work was supported in part by the U.S. Department of Energy under Grant No. DE-FG02-96ER40967.                                              
                                                                               
\section*{References}


\begin{thebibliography}{99}                                                     
                                                                    
\bibitem{hawk}                                                                  
S.W. Hawking, {\it Comm. Math. Phys.} {\bf 43} (1975) 199;                      
G.W. Gibbons and S.W. Hawking {\it Phys. Rev. D}{\bf 15} (1977) 2752.           
\bibitem{hl1}                                                                   
B. Harms and Y. Leblanc, {\it Phys. Rev. D} {\bf 46}, 2334 (1992);              
{\it D} {\bf 47}, 2438 (1993); {\it Ann. Phys.} {\bf 244},                            
262 (1995); {\bf 244}, 272 (1995); {\bf 242}, 265 (1995).                       
\bibitem{chlc}                                                                  
R. Casadio, B. Harms, Y. Leblanc and P.H. Cox, {\it Phys. Rev. D}               
{\bf 55}, 814 (1997); {\it D} {\bf 56}, 4948 (1997).                            
\bibitem{chl}                                                                   
R. Casadio, B. Harms and Y. Leblanc, {\em Statistical mechanics of              
Kerr-Newman dilaton black holes and the bootstrap condition},                   
gr-qc/9706005.                                                                  
\bibitem{mfdbh}                                                                 
R. Casadio, B. Harms and Y. Leblanc, {\it Microfield dynamics of                
black holes}, in preparation.                                                   
\bibitem{fub}                                                                   
S. Fubini, A.J. Hanson and R. Jackiw, {\it Phys. Rev. D} {\bf7},                
1732 (1973);                                                                    
J. Dethlefsen, H.B. Nielsen and H.C. Tze,                                       
{\it Phys. Lett.} {\bf 48B}, 48 (1974);                                         
A. Strumia and G. Venturi, {\it Lett. Nuovo Cimento} {\bf 13},                  
337 (1975).                                                                     
\bibitem{hag}                                                                   
R. Hagedorn, {\it Nuovo Cimento Suppl.} {\bf 3}, 147 (1970);                    
S. Frautschi, {\it Phys. Rev. D} {\bf 3}, 2821 (1971);                          
R.D. Carlitz, {\it Phys. Rev. D} {\bf 5}, 3231 (1972).                          
\bibitem{mfd}                                                                   
Y. Leblanc, in preparation.                                                     
\end{thebibliography}
\end{document}